\documentclass[prb,aps,twocolumn,floatfix,footinbib]{revtex4}
\usepackage{epsfig}
\usepackage{subfigure}
\usepackage{amsfonts}
\usepackage{amsmath}
\usepackage{amssymb}
\usepackage{graphicx}
\begin{document}
\title{Tailoring Chirp in Spin-Lasers}

\author{Guilhem Boeris$^{1,2}$}
\author{Jeongsu Lee$^{1}$}
\author{Karel V\'{y}born\'{y}$^{1,3}$}
\author{Igor \v{Z}uti\'{c}$^{1}$}
\email{zigor@buffalo.edu}
\affiliation{
$^{1}$ Department of Physics, State University of New York at Buffalo, NY 14260, USA \\
$^{2}$ D{\'e}partement de Physique, Ecole Polytechnique, 91128 Palaiseau Cedex, France\\
$^{3}$ Institute of Physics ASCR, v.v.i., Cukrovarnick\'{a} 10, CZ-16253, Praha 6, Czech Republic
}

\date{\today}

\begin{abstract}
The usefulness of semiconductor lasers is often limited by the undesired frequency modulation, 
or  chirp, a direct consequence of the intensity modulation and carrier dependence of the
refractive index in the gain medium. In spin-lasers, realized by injecting,  optically or electrically,  
spin-polarized carriers, we elucidate paths to tailoring chirp. We provide a generalized expression
for chirp in spin-lasers and introduce modulation schemes that could simultaneously eliminate
chirp and enhance the bandwidth, as compared to the conventional (spin-unpolarized) lasers.
\end{abstract}

\maketitle
Many advantages of lasers stem from their modulation response, in which 
refractive index and optical gain depend on carrier density $n$.\cite{Coldren:1995,Yariv:1997}
Modulation  $\delta n(t)$ thus generates both the intensity (photon density)  $\delta S(t)$ and 
frequency modulation $\delta \nu(t)$ of the emitted light. Such $\nu(t)$, known as 
chirp,\cite{Coldren:1995}
is usually a parasitic effect associated with linewidth broadening, enhanced 
dispersion, and limiting the high bit-rate in telecommunication systems.\cite{Petermann:1988} 
Various approaches have therefore focused on low-chirp modulation: 
pulse shaping,\cite{Petermann:1988} injection locking,\cite{vanTartwijk:1998} 
temperature modulation,\cite{Gorfinkel1993:APL} 
and employing quantum dots as the gain region.\cite{Chuang:2009}
In conventional lasers 
for small signal analysis\cite{Chuang:2009} (SSA) in which the quantities of interest 
are decomposed into a steady state and modulated part $X=X_0+\delta X(t)$, 
the chirp is given by\cite{Coldren:1995} 
\begin{equation}
\delta \nu(t)
=[\Gamma g_0/(4\pi)]\, \alpha_0 \,\delta n(t), 
\label{eq:chirp}
\end{equation}
where $\Gamma$ is the optical confinement factor , $g_0$ the gain coefficient,
and   $\alpha_{0}=(\partial \hat{n}_r/\partial n)/(\partial \hat{n}_i/\partial n)$
is the linewidth enhancement factor,\cite{Chuang:2009} expressed in terms of complex
refraction index $\hat{n} = \hat{n}_r+i \hat{n}_i$ in the active region. 

\begin{figure}[tbp] 
\centering
\includegraphics[height=1.14in]{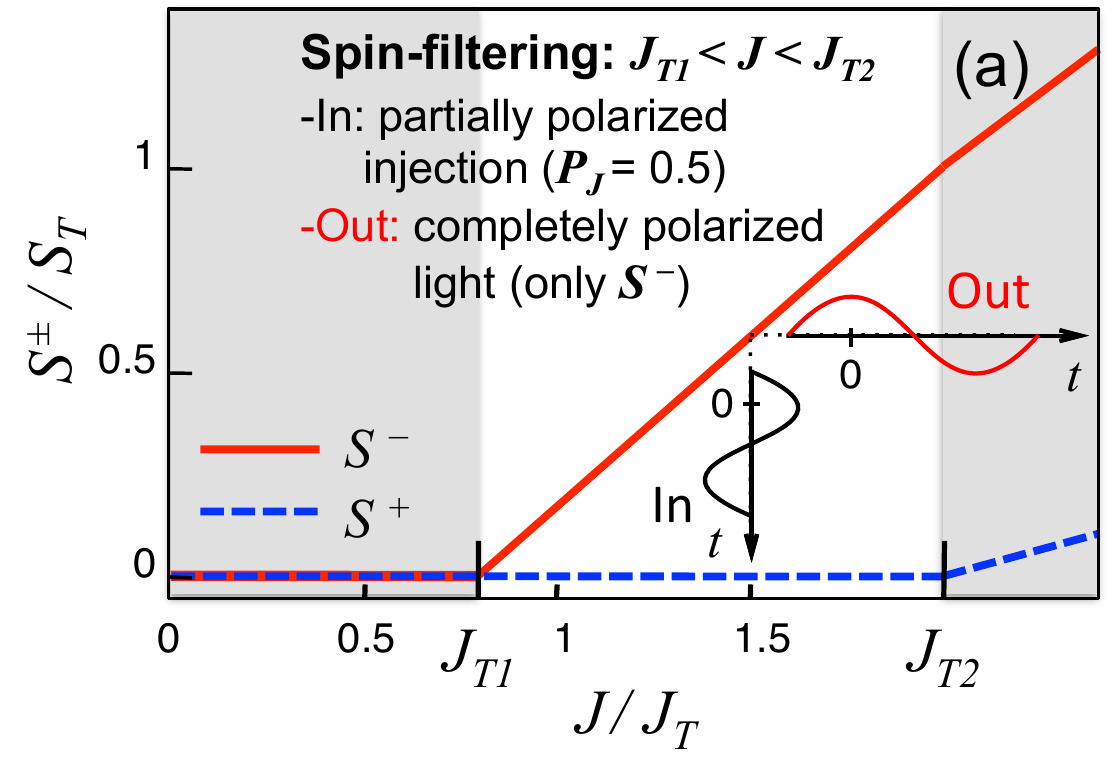}
\includegraphics[height=1.14in]{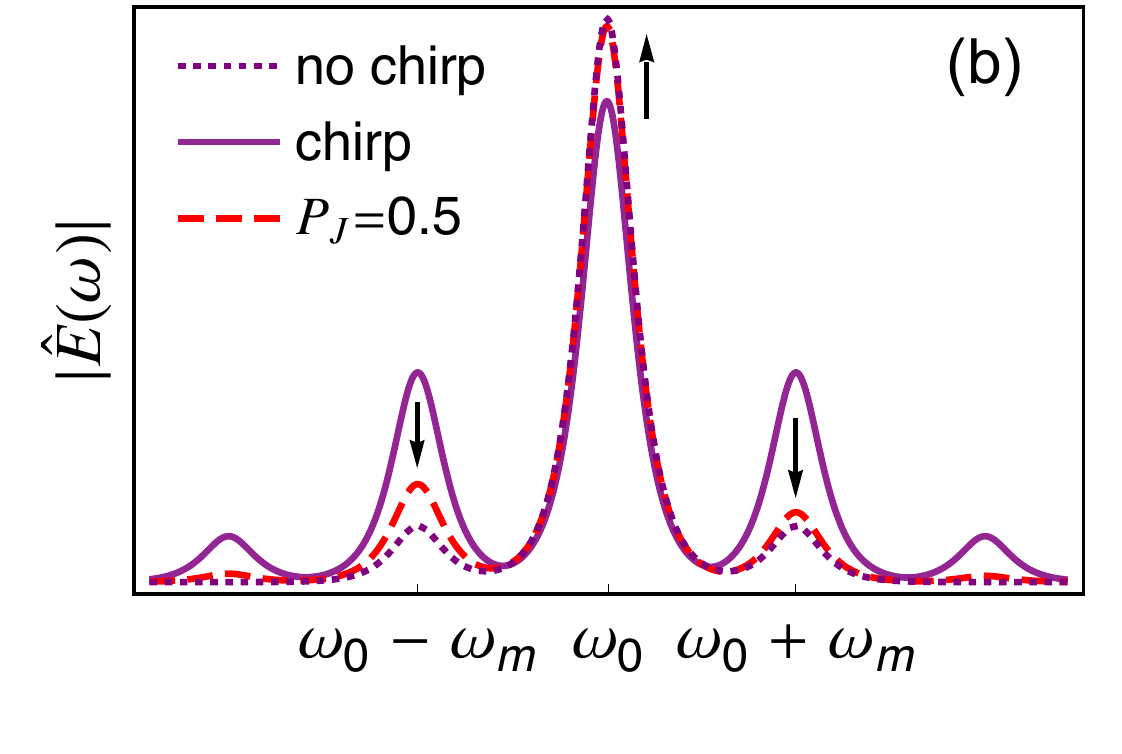}
\caption{(Color online)
(a) Helicity-resolved photon density ($S^\pm$) as a function of injection ($J$),
normalized to $S_T=S(2J_T)$  and unpolarized injection threshold $J_T$, respectively.
For spin-polarized injection, $|P_J|>0$, 
there are two thresholds\cite{Lee2010:APL} 
$J_{T1,2}$ for $S^\mp$. AM (harmonic curves) for $J \in (J_{T1},J_{T2})$
yields modulation of fully polarized light (spin-filtering, unshaded area). 
(b) Broadened electric field spectrum for AM. Conventional lasers ($P_J=0$) without 
(dotted line) and with chirp (solid), and spin-laser with $P_J=0.5$ (dashed) are shown. 
Arrows indicate the chirp reduction by spin injection. Modulation amplitudes for $P_J=0.5$ 
and $P_J= 0$ are chosen to provide the same spectra for no chirp. The choice
of colors reflects that an unpolarized $S$ is an equal weight superposition of  
$S^+$ and $S^-$, while for $P_J=0.5$ the emitted light is $S^-$.  
}
\label{fig:01}
\end{figure}

In the emerging class of semiconductor lasers, known as spin-lasers,%
\cite{Hallstein1997:PRB, Rudolph2003:APL,Rudolph2005:APL,Holub2007:PRL,Hovel2008:APL,Saha2008:APL,%
Fujino2009:APL,Gothgen2008:APL,Vurgaftman2008:APL,Lee2010:APL,%
Saha2010:PRB, Zutic2011:CRC,Iba2001:APL,Al-Seyab2011:PJ,Holub2011:PRB, Banerjee2011:JAP, Lee2012:PRB} 
with total injection $J=J_+ - J_-$ containing inequivalent spin up/down contributions
($J_+$, $J_-$), we expect additional possibilities for tailoring chirp. 
$J_+ \neq J_-$ is realized using circularly-polarized photoexcitation or electrical injection 
from a magnetic contact.  The polarization of   
emitted light resolved in two helicities, $S=S^+ + S^-$,
can be understood from the optical selection rules.\cite{Zutic2004:RMP} 
For example, in the quantum well-based  spin-lasers with $J$ close 
to the lasing threshold, recombination 
of spin-up (spin-down) electrons and heavy holes yields $S^-$ ($S^+$) polarized light.
Both amplitude modulation (AM) $\delta J(t)$ [see Fig.~\ref{fig:01}(a)]
and polarization modulation (PM) $\delta P_J(t)$, of  injection polarization\cite{Zutic2004:RMP} 
 $P_J=(J_+ - J_-)/(J_++J_-)$, can be readily implemented. With PM the emitted
light could be modulated even at a {\em fixed} $J$ and $n$.\cite{Lee2010:APL}
While Eq.~(\ref{eq:chirp}) then suggests a chirp-free operation, we show that 
such a simple reasoning is not always true and suitable generalization 
for chirp in spin-lasers is required. 

Our generalized picture reveals  that AM and PM in spin-lasers
enable both reduced chirp and enhanced modulation bandwidth, 
as compared to their spin-unpolarized ($P_J=0$) counterparts. 
PM could also provide an efficient spin communication.\cite{Dery2011:APL}

The chirp can be simply quantified by comparing the ratio
of the central and first sideband peaks in the emitted light.\cite{Arakawa1985:APL} 
To visualize
this effect, in Fig.~\ref{fig:01}(b) we show the spectrum of electric field
which can be written as\cite{Yariv:1997}
\begin{equation}
E(t)\simeq E_0[1+\delta S(t)/(2S_0)] \, \mathrm{Re} \{e^{i [2\pi \nu_0 t + \phi (t)]} \}, 
\label{eq:field}
\end{equation}
where $E_0$ is a real amplitude of the field, 
the phase is $\phi(t)=2 \pi \int_0^t \delta \nu(t') dt'$, and 
$\nu_0$ ($\omega_0$) is (angular) frequency of the output light. 
Using rate equations (REs) we calculate harmonic modulation with $\omega_m$ in SSA\cite{REs} 
and  obtain $\phi(t)=[ | \delta \nu (\omega_m ) |/\nu_m ] \sin (\omega_m t + \phi_\nu)$, 
where $\phi_{\nu}=\mathrm{arg}[\delta \nu(\omega_m)]$. 
The undesirable alteration to the original spectrum caused by chirp can be quantified 
by the ratio between the heights of the first sidebands 
with and without chirp. For spin-unpolarized lasers, an indentity\cite{Yariv:1997}
$e^{i\delta \sin x}=\sum_{n=-\infty}^{\infty} J_n(\delta)e^{inx}$, with 
asymptotic approximation $\delta\ll1$ for Bessel functions $J_n(\delta)$, 
leads to\cite{Coldren:1995,Petermann:1988}
\begin{equation}
\frac{\mathrm{sideband\, height\,with\,chirp}}{\mathrm{sideband\,  height\,  without\, chirp}}
 \simeq \sqrt{1 + 4 \bigg(\frac{\mathrm{FM}}{\mathrm{IM}} \bigg)^2},
\label{eq:sidepeak}
\end{equation}
where the ratio of  frequency and intensity modulation index FM/IM can 
be expressed as\cite{Coldren:1995,Petermann:1988} 
\begin{equation}
\mathrm{FM}/\mathrm{IM} =[\delta \nu(\omega_m) /\nu_m]/[\delta S(\omega_m)/S_0]. 
\label{eq:FM/IMdef}
\end{equation}
Equation~(\ref{eq:sidepeak})  accurately gives the variation of the first sidebands 
in Fig.~\ref{fig:01}(b). The phase induced by the chirp also creates higher order 
sidebands further away. 
However, by the spin-polarized injection modulation, 
chirp and thus alteration of the spectrum can be suppressed. 

To define chirp in spin-lasers, we recall that the generalization of the 
usual model of optical gain term\cite{Gothgen2008:APL, Lee2010:APL} 
is $g_0(n-n_\text{tran}) \to 
g_0(n_\pm + p_\pm - n_\text{tran}) = g_0[(3/2)n_\pm + (1/2) n_\mp - n_\text{tran}]$, 
where $g_0$ is density-independent coefficient, 
$n_\text{tran}$ the transparency density, and $n_\pm$ $(p_\pm)$ are electron (hole)
spin-resolved density. 
Here 3:1 ratio of $n_\pm$ contributions
follows from the charge neutrality and  the very fast spin relaxation 
of holes\cite{Gothgen2008:APL} $p_\pm=n/2$
and this ratio reflects also the gain anisotropy for $S^+$ and $S^-$.

For spin-lasers  the generalization of Eq.~(\ref{eq:chirp}) is then
\begin{equation}
\delta \nu(t) =\frac{\Gamma g_0}{4\pi} \left[
\frac{3}{2}\alpha_+ \delta n_+(t)+\frac{1}{2}\alpha_- \delta n_-(t) \right],
\label{eq:spinchirp}
\end{equation}
where we  focus  on the spin-filtering regime [Fig.~\ref{fig:01}(a)],
 $J\in (J_{T1}, J_{T2})$, and 
 $\alpha_{\pm}=(\partial \hat{n}_r/\partial n_{\pm})/(\partial \hat{n}_i/\partial n_{\pm})$.\cite{alpha}
For $P_J>0$ the spin filtering implies $S^-$ emitted light.\cite{chirp}
When $P_J=0$  (thus $n_+=n_-$), Eq.~(\ref{eq:spinchirp}) 
reduces to Eq.~(\ref{eq:chirp}) since 
\begin{equation}
\alpha_0 = (3 \alpha_+ + \alpha_-)/4.
\label{eq:alphas}
\end{equation}
While  for $P_J\neq0$, Eq.~(\ref{eq:alphas}) is not always true (since $\alpha_\pm$
depends on $n_\pm$), we still use it to relate  $\alpha_\pm$ and $\alpha_0$. 
This approximation is  precise  for $J$ slightly below $J_{T2}=J_T/(1-P_{J0})$ where  
$n_+$$-$$n_- \to 0$. 

With typical spin-lasers, realized as vertical cavity surface emitting 
lasers,\cite{Rudolph2003:APL,Rudolph2005:APL,Holub2007:PRL,Hovel2008:APL,%
Fujino2009:APL,Saha2010:PRB} 
in the spin-filtering regime it is accurate  to use\cite{Gothgen2008:APL}
vanishing gain compression and spontaneous emission factors 
($\epsilon=\beta=0$).\cite{beta}
With a generalized chirp formulation  [Eq.~(\ref{eq:spinchirp})], we employ REs\cite{Lee2010:APL}
and SSA to obtain the results from Fig.~\ref{fig:01}(b). We confirm the chirp 
suppression in spin-lasers  with the spectrum approaching the chirp-free case.

In conventional lasers, the chirp  reduction is particularly important for high-frequency 
modulation where the transient chirp [$\propto d \ln S(t)/dt$, only weakly $\epsilon$-dependent]
is the dominant contribution.\cite{Coldren:1995,Yariv:1997,Petermann:1988}
FM/IM is a constant
\begin{equation}
\frac{\delta\nu(\omega_m)/\nu_m}{\delta S(\omega_m)/S_0} = - i\frac{\alpha_0}{2}\, ,
\label{eq:FM/IM}
\end{equation}
which provides both a suitable way to experimentally extract\cite{Coldren:1995} the linewidth 
enhancement factor  $\alpha_0$,  and a simple comparison for chirp in spin-lasers.
In the spin-filtering regime 
$|\text{FM/IM}|$  depends on the modulation frequency $\omega_m$ and
the ratio $\rho\equiv\alpha_+/\alpha_-$ [see Eq.~(\ref{eq:spinchirp})]
\begin{equation}
\bigg| \frac{\delta\nu/\nu_m}{\delta S/S_0} \bigg| / \frac{\alpha_0}{2} = 
\frac{3 \rho \delta n_+(\omega_m) + \delta n_-(\omega_m)}{3 \delta n_+(\omega_m) + 
\delta n_-(\omega_m)} \bigg( \frac{ 4}{1 + 3 \rho} \bigg). 
\label{eq:spinFM/IM}
\end{equation}

\begin{figure}[tbp] 
\centering
\includegraphics[height=2.28in]{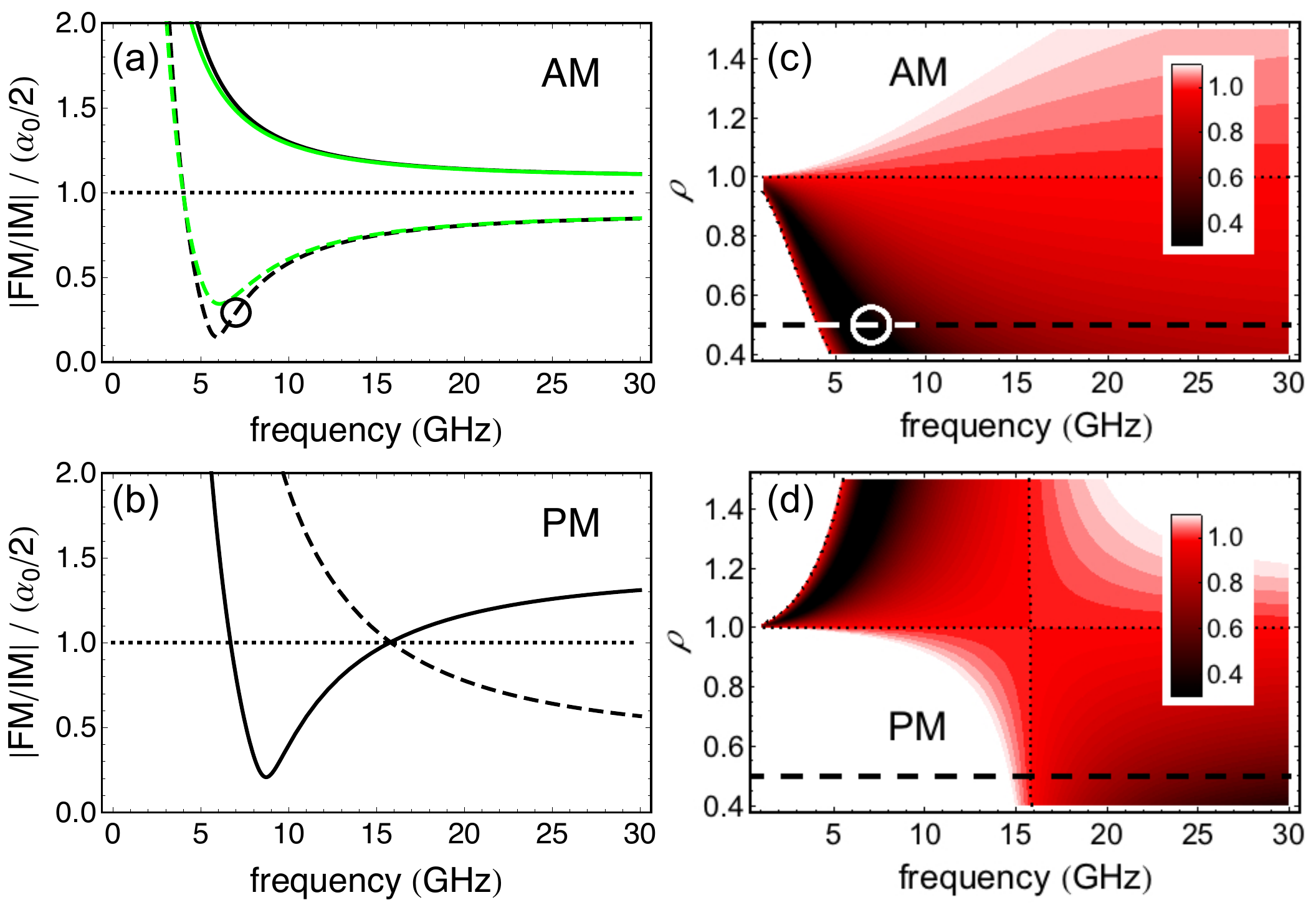}
\caption{(Color online) 
 $|\text{FM/IM}|$ normalized to the conventional value $\alpha_0/2$ 
for (a) AM  and (b) PM, shown for  $\rho\equiv \alpha_+/\alpha_- = 2$ 
(solid) and  $\rho = 0.5$ (dashed). Green (gray) curves reveal only a small
change for finite electron spin relaxation time,\cite{REs} $\tau_s=\tau_r$.
The regime of reduced chirp 
in spin-lasers (darker regions) is delimited 
with dotted lines for (c) AM and (d) PM. The circle in (a) and (c) for $\rho=0.5$
represents the sampling point to generate Fig.~\ref{fig:01}(b). 
$J_0 = 1.9  J_T$ and  $P_{J0} = 0.5$ are used  in (a)-(d).
}
\label{fig:02}
\end{figure}

$|\text{FM/IM}|$ of spin lasers is shown in Fig.~\ref{fig:02}.
A choice of $\rho \in [0.5,2]$
is motivated by our preliminary microscopic calculation
(Kubo formalism)  of $\alpha_+$ and $\alpha_-$ for GaAs. 
The normalized ratio $|\text{FM/IM}| < 1$ represents the chirp reduction relative
to conventional lasers. 
For AM a change  $\rho=2 \to 0.5$ leads to a smaller chirp for
all range of modulation frequencies in Fig.~\ref{fig:02}(a). Black
and gray (green) curves show only a small change in the results 
for electron spin relaxation time $\tau_s$, being infinite and equal 
to the recombination time $\tau_r$,  respectively. 
Since in spin-lasers at 300 K $\tau_s/\tau_r \sim 10$,\cite{Fujino2009:APL} 
it is accurate to choose $\tau_s \to \infty$ in REs for the rest of our analysis. 

For PM  in Fig.~\ref{fig:02}(b) the same change $\rho=2 \to 0.5$ 
yields a non-monotonic effect on the chirp reduction which, compared 
to the conventional lasers, is realized at 
$\nu_m \lesssim 16$ GHz ($\rho=2$ ) and at $\nu_m \gtrsim 16$ GHz 
($\rho=0.5$), respectively. These trends for AM and PM are 
further shown in Figs.~\ref{fig:02}(c) and (d)  for a range of $\rho$, where the 
region of the favorable $|\text{FM/IM}|$ reduction is delimited with dotted lines. 
Consistent with  Eq.~(\ref{eq:spinFM/IM}), $|\text{FM/IM}|$ at $\rho=1$  yields  the conventional  
value $\alpha_0/2$, for both AM and PM. 
Since such a
conventional value is retained even for PM and $\delta n(t)=0$,  there is a striking
difference between the usual chirp in  Eq.~(\ref{eq:chirp}), and that for spin-lasers in Eq.~(\ref{eq:spinchirp}). 
 
Our discussion of  FM/IM shows that the chirp is not 
completely removed  using PM or AM. However, it  is possible to achieve zero-chirp 
by introducing a scheme we term complex modulation (CM): one of the
spin-resolved injections ($J_+$ for $P_{J0} >0$ ) is the input signal, while the
the other is used only to cancel the chirp.
From Eq.~(\ref{eq:spinFM/IM}),  the zero-chirp condition is 
$\delta n_-(\omega_m)/\delta n_+(\omega_m) = -3 \alpha_+/\alpha_-=-3\rho$, 
which can be satisfied by introducing a 
chirp-tailoring function $\kappa(\omega_m)$  obtained from SSA 
\begin{equation}
\delta J_-(\omega_m) = \kappa(\omega_m)\, \delta J_+(\omega_m).
\label{eq:CM}
\end{equation}
Here $\delta J_+$ is the input modulation responsible for the modulation of emitted 
light $\delta S^-$, while the correction current $\delta J_-$ 
compensates the variation of the carrier density to reduce the chirp. 
\begin{figure}[tbp] 
\centering
\includegraphics[width=3.42in]{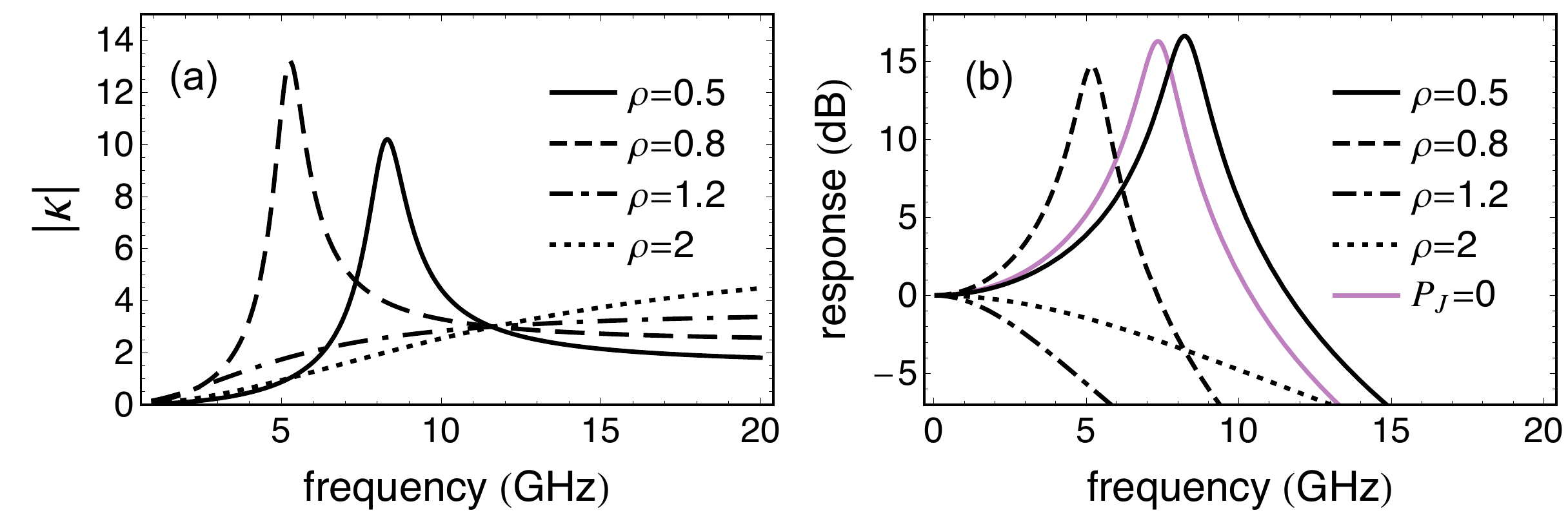} 
 \caption{(Color online) 
SSA of CM. 
(a) Chirp-tailoring function $\kappa(\nu_m)$  and 
(b) modulation response $|R(\nu_m)/R(0)|^2$ are shown for 
$J_0 = 1.2 \, J_T$, $P_{J0} = 0.5$ and different $\rho$'s.
The response of conventional laser (AM, $P_J=0$) is given (gray/purple) for comparison. 
}
\label{fig:03}
\end{figure}

We next use SSA to consider the implications of CM on the modulation bandwidth, 
shown together with the chirp-tailoring function $\kappa$ in Fig.~\ref{fig:03}.  
The CM relaxation oscillation frequency, represented by the peak positions  
in Figs.~\ref{fig:03}(a) and (b) for $\rho\leq1$,  can be expressed as
\begin{eqnarray}
\omega_R^{CM}\simeq\{\Gamma g_0 J_T(J/J_{T1}-1) (1-\rho)\}^{1/2}\, ,
\label{eq:CM}
\end{eqnarray}
where $J_{T1}=J_T/(1+P_{J0}/2)$ is the reduced threshold in a spin-laser.\cite{Gothgen2008:APL} 
The peak positions coincide  for $|\kappa(\omega_m)|$ 
and for the  modulation response function\cite{Lee2010:APL} 
$R(\omega_m)=|\delta S^-(\omega_m)/\delta J_+(\omega_m)|$  
because the character of 
$J_-(\omega_m) \propto \kappa(\omega_m)$ propagates 
through $n_\pm$ and $S^-$ 
into $R(\omega_m)$.
For $\rho>1$, $|\kappa(\omega_m)|$ increases monotonically with $\omega_m$ showing 
no peak. Zero-chirp is  not feasible for $\rho=1$ since
it is the same FM/IM as in conventional lasers [Eq.~(\ref{eq:spinFM/IM})]. 

By comparing $\omega_R^{CM}$ in Eq.~(\ref{eq:CM})  for $\rho \leq 1$ to 
$\omega_R^{AM}$ and $\omega_R^{PM}$ from Ref.~\onlinecite{Lee2010:APL}:
$\omega_R^{AM,PM}\simeq\{\Gamma g_0 J_T(J/J_{T1}-1)\}^{1/2}$, 
we see that CM has narrower bandwidth than AM and PM (estimated by 
$\omega_R$), for the same $P_{J0}$. While CM provides a path for removing chirp, 
it may come at the cost of a reduced bandwidth. 
However, an optimized value of  $\rho= 0.5$ in  Fig.~\ref{fig:03}(b) yields simultaneously 
zero chirp and bandwidth enhancement, as compared to conventional lasers.

What about experimental  feasibility to control  chirp in spin-lasers?
While CM has yet to be attempted, it can be viewed as  a combination
of AM and PM which individually already  lead to an improved chirp 
(Figs.~\ref{fig:01} and \ref{fig:02}) and have 
been demonstrated in spin-lasers. 
Slow PM has been realized\cite{Rudolph2003:APL} using a Soleil-Babinet
polarization retarder at a fixed $J \in (J_{T1}, J_{T2})$. 
Fast PM ($\nu_m \sim 40$ GHz) can be  implemented with a
coherent electron spin precession in a transverse 
magnetic field,\cite{Hallstein1997:PRB} 
or a mode conversion in an electro-optic modulator.\cite{Bull2004:PSPIE} 
Recent advances in using  birefringence for PM\cite{Gerhardt2011:APL} 
suggest that chirp reduction in spin-lasers could be feasible  
at higher injection, beyond the spin-filtering regime we have considered.

To further enhance the opportunities in spin-lasers, it would be helpful to 
utilize other gain media and achieve technologically important
emission at $1.3$ and $1.55$ $\mu$m. We expect that our
proposals will stimulate additional work towards understanding
the spin-dependence of refractive index  (already used for
fast all-optical switching\cite{Nishikawa1995:APL}) 
and its implications for spin-lasers.

We thank H. Dery, R. Oszwa{\l}dowski, A. Petrou, and N. Tesa\v rov\'a 
for discussions. This work was supported by the NSF-ECCS, 
AFOSR-DCT, U.S. ONR,  NSF-NRI NEB 2020, and SRC.


\begin{thebibliography}{99}
\bibitem{chirp}
If there is an emitted light with the other helicity, the chirp from $S^+$ signal, $\delta \nu'(t),$ can be written
analogously: 
$\delta \nu'(t) =\Gamma g_0/(4\pi)[(1/2)\alpha'_+ \delta n_+(t)+(3/2)\alpha'_- \delta n_-(t)]$.
\end{thebibliography}
\end{document}